\def\year{2021}\relax
\documentclass[letterpaper]{article} 
\usepackage{aaai21}  
\usepackage{times}  
\usepackage{helvet} 
\usepackage{courier}  
\usepackage[hyphens]{url}  
\usepackage{graphicx} 
\usepackage{natbib}  
\usepackage{caption} 
\usepackage{color}
\title{CoVaxxy: A Collection of English-Language\\Twitter Posts about COVID-19 Vaccines}

\author{Matthew R. DeVerna,$^*$\textsuperscript{\rm 1} Francesco Pierri,$^*$\textsuperscript{\rm 1,2}
Bao Tran Truong,$^*$\textsuperscript{\rm 1} John Bollenbacher,$^*$\textsuperscript{\rm 1}\\
David Axelrod,\textsuperscript{\rm 1} Niklas Loynes,\textsuperscript{\rm 3,4} Christopher Torres-Lugo,\textsuperscript{\rm 1} Kai-Cheng Yang,\textsuperscript{\rm 1}\\ Filippo Menczer,$^\dagger$\textsuperscript{\rm 1} and John Bryden$^\dagger$\textsuperscript{\rm 1}
\\}

\affiliations{
    \textsuperscript{\rm 1}Observatory on Social Media, Indiana University, Bloomington, USA\\
    \textsuperscript{\rm 2}Dipartimento di Elettronica, Informazione e Bioingegneria, Politecnico di Milano, Italy\\
    \textsuperscript{\rm 3}School of Social Sciences, University of Manchester, UK\\
    \textsuperscript{\rm 4}Corridor Labs, London, UK\\
    \{mdeverna, fpierri, baotruon, jmbollen, daaxelro, nlyones, torresch, yangkc, fil, jabryden\}@iu.edu\\
}

\newcommand{\keyword}[1]{{\texttt{#1}}}

\begin{document}
\maketitle

\def\thefootnote{*}\footnotetext{These authors contributed equally to this work}

\def\thefootnote{$\dagger$}\footnotetext{These authors contributed equally to this work}

\def\thefootnote{\arabic{footnote}}

\setcounter{footnote}{0}

\begin{abstract}
With a substantial proportion of the population currently hesitant to take the COVID-19 vaccine, it is important that people have access to accurate information. 
However, there is a large amount of low-credibility information about vaccines spreading on social media. 
In this paper, we present the \textit{CoVaxxy} dataset, a growing collection of English-language Twitter posts about COVID-19 vaccines.
Using one week of data, we provide statistics regarding the numbers of tweets over time, the hashtags used, and the websites shared. 
We also illustrate how these data might be utilized by performing an analysis of the prevalence over time of high- and low-credibility sources, topic groups of hashtags, and geographical distributions. 
Additionally, we develop and present the \textit{CoVaxxy} dashboard, allowing people to visualize the relationship between COVID-19 vaccine adoption and U.S. geo-located posts in our dataset.
This dataset can be used to study the impact of online information on COVID-19 health outcomes (e.g., vaccine uptake) and our dashboard can help with exploration of the data.
\end{abstract}

\section{Introduction}

The COVID-19 pandemic has killed two million people and infected 93 million around the world as of mid-January, 2021~\cite{Dong_Du_Gardner_2020}.
Vaccines will be critical in our fight to end the COVID-19 pandemic~\cite{orenstein2017simply}.
It is estimated that around 60-70\% of the population will need to be vaccinated against COVID-19 to achieve herd immunity~\cite{Aguas2020threshold}.
However, recent surveys have found that only 40-60\% of American adults reported that they would take a COVID-19 vaccine~\cite{pew2020intent, KFF2020intent}.
With these currently predicted levels of \emph{vaccine hesitancy}, it is unlikely we will reach herd immunity; COVID-19 will remain endemic.

A possible driver for vaccine hesitancy is the anti-vaccination movement.
This movement has been on the rise in the U.S. for two decades, beginning with unfounded fears over a Measles, Mumps and Rubella (MMR) vaccine~\cite{hussainAntivaccinationMovementRegression}. 
The vocal online presence of the anti-vaccination movement has undermined confidence in vaccines. 
Worse, resistance to the COVID-19 vaccines is currently much more prevalent than resistance to the MMR vaccine.
Since COVID-19 vaccine hesitancy and its drivers remain understudied, a goal of our project is to help address this gap. 

There is a growing body of evidence linking social media and the anti-vaccination movement to vaccine hesitancy~\cite{broniatowski2018weaponized,burkiVaccineMisinformationSocial2019,johnsonOnlineCompetitionPro2020b}. Studies show that vaccine hesitancy in one's peer group is associated with future hesitancy~\cite{brunsonImpactSocialNetworks2013}, and that misinformation spread on social networks is linked to poor compliance with public health guidance about COVID-19~\cite{Roozenbeek2020}. Based on these findings, the core hypothesis behind this project is that the social spread of vaccine misinformation and vaccine hesitancy will impact public health outcomes such as vaccine uptake and COVID-19 mortality rate.  

Here we present a collection of English-language posts related to the COVID-19 vaccines on Twitter.  
The collection is exempt from IRB review as it only includes tweet IDs of public messages. 
This allows us to comply with the Twitter Terms of Service while making the data available to both researchers and the general public. 
Although there has been previous work presenting COVID-19 Twitter datasets~\cite{chen2020tracking, Huang2020coviddata, lamsal2020coviddata}, our work focuses specifically on discussion of COVID-19 vaccines and related public health outcomes.

The \textit{CoVaxxy} dataset will enable researchers to study vaccine misinformation and hesitancy, and their relationship to public health outcomes.  We will use established techniques to track vaccine misinformation within the data, along with misinformation superspreaders, coordinated campaigns, and automated accounts~\cite{yang2019bot,yang2020covid, Pierri2020epj, Pierri2020scirep,pacheco2020uncovering}. We will also relate this social media data to geographic public health data (such as COVID-19 mortality and vaccine uptake rates) by using geolocation data within the dataset. 

In this paper we describe the methods used to create the \textit{CoVaxxy} dataset. Using one week of data, we provide a descriptive analysis and illustrate how our data could be used to answer various research questions. We also present the \textit{CoVaxxy Dashboard}, a tool intended for the public to track key insights drawn from the data. Opportunities and limitations are discussed as we draw conclusions.

\section{Dataset Curation}

Our key data collection goal is to download a complete set of Twitter posts related to COVID-19 vaccines. In this section we describe our methodology for selecting appropriate keywords to achieve such a coverage.
We then describe our architecture with server redundancy to maintain an unbroken stream of Twitter data containing these keywords. 

\subsection{Identifying COVID-19 Vaccines Content}

To create as complete a set of Twitter posts related to COVID-19 vaccines as possible, we carefully select a list of keywords through a snowball sampling technique~\cite{conover2012,yang2019bot}.
We start with the two most relevant keywords, i.e.,  \keyword{covid} and \keyword{vaccine}, as our initial seeds.
Keywords also match hashtags, URLs, and substrings. For example, \keyword{covid} matches ``cnn.com/covid'' and ``\#covid.''
Next, we gather tweets utilizing the filtered stream endpoint of the Twitter API\footnote{\url{https://developer.twitter.com/en/docs/twitter-api/v1/tweets/filter-realtime/overview}} for three hours.
From these gathered tweets, we then identify potential keywords that frequently co-occur with the seeds. These keywords are separately reviewed by two authors and added to the seed list if both agree that a keyword is related to our topic.
This process was repeated six times between Dec. 15, 2020 and Jan. 2, 2021 with each iteration's data collection taking place at different times of the day to capture tweets from different geographic areas and demographics. The seed list serves as our initial keyword list.

We further refine the keyword list by manually combining certain keywords into composites, leveraging the query syntax of Twitter's filtered stream API.
For example, using \keyword{covid19 pfizer} as a single composite matching phrase will capture tweets that contain \textit{both} ``covid19'' \textit{and} ``pfizer.''
On the other hand, including \keyword{covid19} and \keyword{pfizer} as separate keywords will capture tweets that contain ``covid19'' \textit{or} ``pfizer,'' which we consider as too broad for our analysis.
The final keyword list includes 76 (single or composite) keywords. 
Constructing various composites of relevant keywords in this way ensures the dataset is broad enough to include most relevant conversations while excluding tweets that are not related to the vaccine discussion.

\subsection{Content Coverage}

To demonstrate the effectiveness of the snowball sampling technique introduced above, we calculate the popularity of each keyword in the final list by the number of unique tweets and unique users associated with it.

\begin{figure}
    \centering
    \includegraphics[width=\columnwidth]{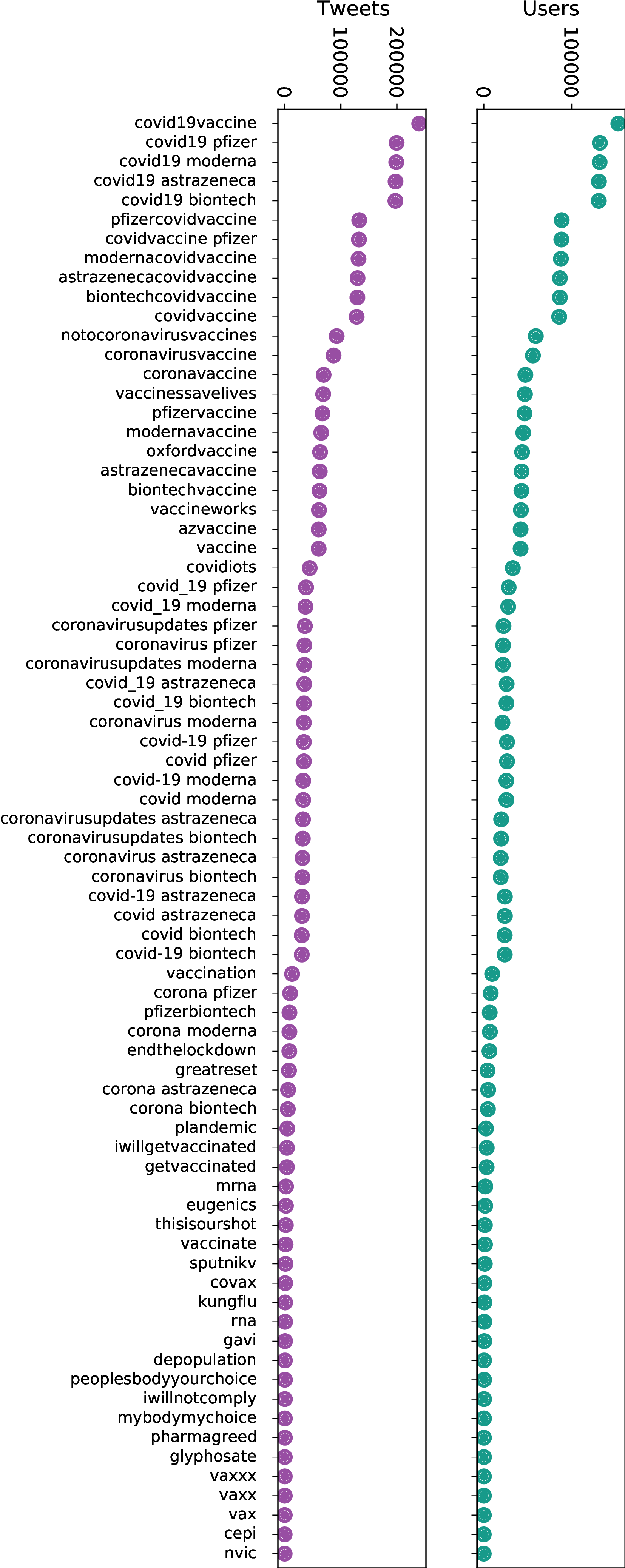}
    \caption{Number of tweets (purple, left) and users (green, right) captured by each keyword/phrase in the final list (ranked by popularity) between January 4--11, 2021.
}
    \label{fig:ht_diminishreturn}
\end{figure}

Figure~\ref{fig:ht_diminishreturn}, where keywords are ranked by popularity, shows that additional keywords beyond the 60 most popular ones tend to capture very small numbers of users and tweets, relative to other keywords in the collection. This suggests that including more keywords in the seed list described above is not likely to alter the size and structure of the dataset significantly. In fact, the inclusion of additional keywords could be redundant, due to the co-occurrence of multiple keywords and hashtags in a single tweet, especially for the most popular terms. Thus, we believe that our set of keywords provides reasonable coverage and is representative of tweets communicating about COVID-19 vaccines.  

As the collection of tweets is intended to persist over time, new relevant keywords may emerge.
To ensure that the keyword list remains comprehensive throughout the entire data collection period, our team will continue to monitor the ongoing public discussion related to COVID-19 vaccinations and update the list with important emerging keywords, if necessary.

\section{CoVaxxy Infrastructure}

\subsection{Data Collection Architecture}

\begin{figure*}
    \centering
    \includegraphics[width=1.5\columnwidth]{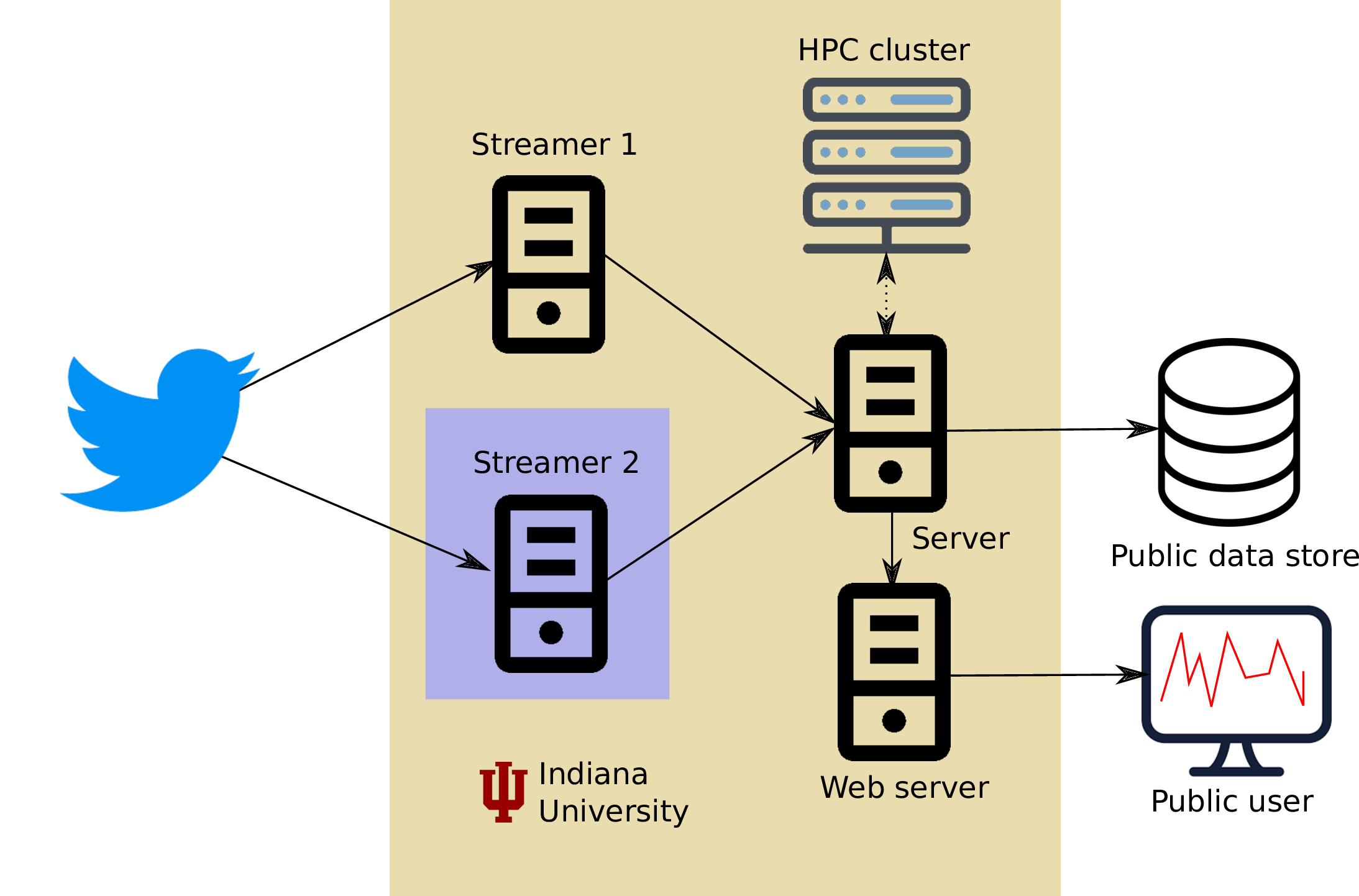}
    \caption{The VM server architecture for the \textit{CoVaxxy} project. Data flows in the direction of the arrows. Machines in the larger yellow box are hosted by Indiana University. The VM ``Streamer 2,'' in the embedded blue box, is hosted by the Texas Advanced Computing Center.}
    \label{fig:ht_serverarchtecture}
\end{figure*}

Our server architecture (Figure~\ref{fig:ht_serverarchtecture}) is designed to collect and process large quantities of data.
This infrastructure is hosted by Extreme Science and Engineering Discovery Environment (XSEDE) Jetstream virtual machines (VMs)~\cite{xsede,stewartJetstreamSelfprovisionedScalable2015}.
To maintain the integrity of our tweet streaming pipeline, we have incorporated redundancy.
We maintain two \textit{streamer} (stream collection) VMs in different U.S. states so that if one suffers a fault we can use data from the other. These servers connect to Twitter's filtered stream API to collect tweets that match any of the keywords in real time. We use the language metadata to filter out non-English tweets. 

The data from the two streamers is collated on a general purpose server VM where we run data analysis.
The server VM is also linked to Indiana University's high performance computing infrastructure for running advanced analyses.

We upload new data files to a public data repository~\cite{covaxxy_dataset} each day\footnote{\url{https://doi.org/10.5281/zenodo.4526494}} and will continue to do so as long as the topic of COVID-19 vaccinations remains relevant in public discourse.
This repository also includes our list of keywords.
In compliance with Twitter's Terms, we are only able to share tweet IDs with the public.
One can re-hydrate the dataset by querying the Twitter API or using tools like Hydrator\footnote{\url{https://github.com/DocNow/hydrator}} or twarc\footnote{\url{https://github.com/DocNow/twarc}}.

Finally, a web server provides access to the data on the server VM through the interactive \textit{CoVaxxy} dashboard, described next.

\subsection{Dashboard}

Existing COVID-19 visualization tools include those by Johns Hopkins University~\cite{Dong_Du_Gardner_2020} and The Atlantic.\footnote{\url{https://covidtracking.com/}} These trackers address hospitalization and mortality. 
Another dashboard from the Fondazione Bruno Kessler reports on the proportions of misinformation and epidemic-related statistics (e.g., confirmed cases and deaths) per country.\footnote{\url{https://covid19obs.fbk.eu}} 
Finally, the Our World in Data COVID-19 vaccination dataset publishes vaccine uptake information by country.\footnote{\url{https://ourworldindata.org/covid-vaccinations}}

\begin{figure*}
    \centering
    \includegraphics[width=\textwidth]{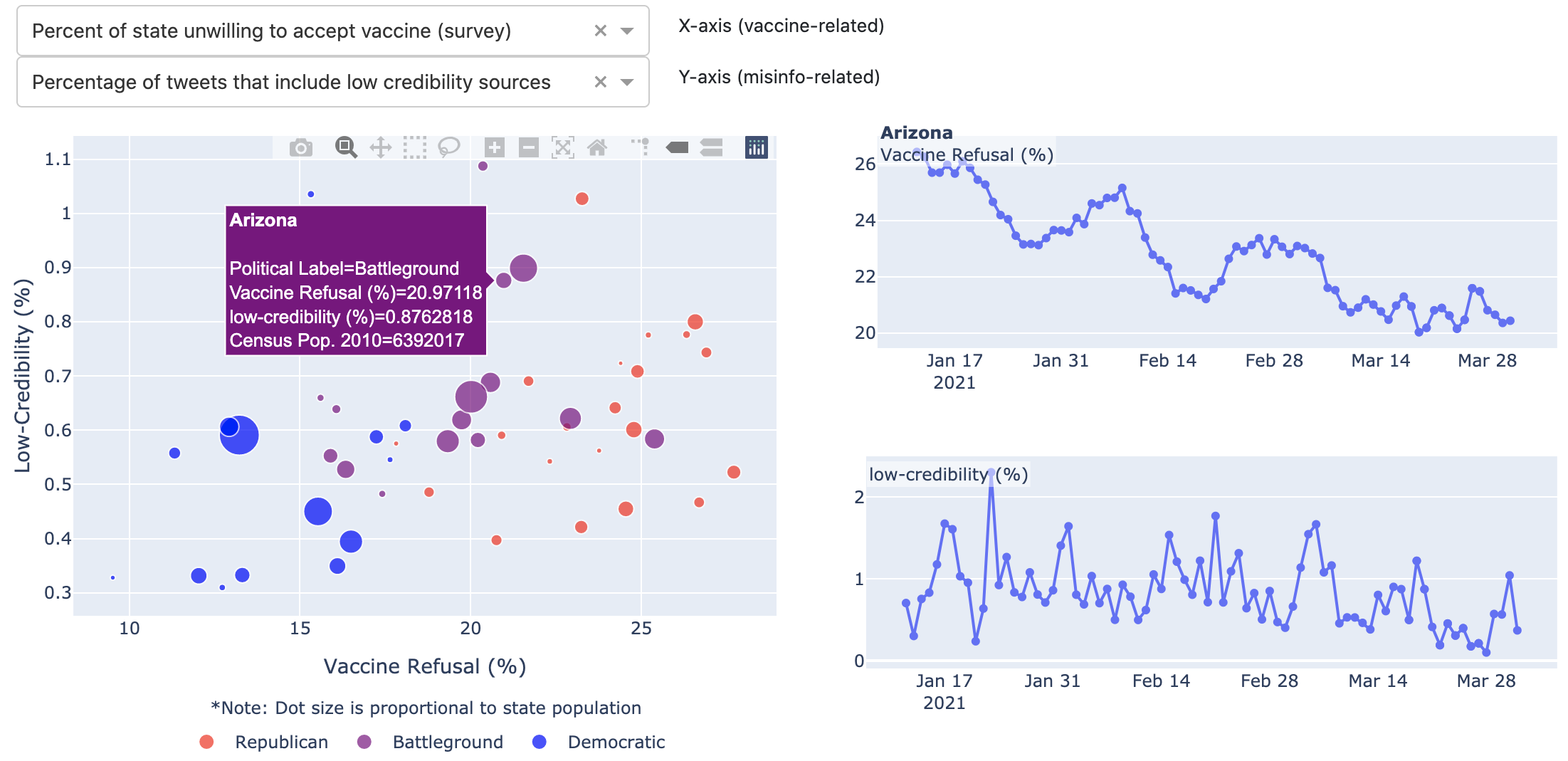}
    \caption{Example visualization from the \textit{CoVaxxy} web dashboard. This visualization lets users plot relationships (at the state-level) between vaccine-related and misinformation-related data. The left figure's axes are selected from the dropdowns, displaying the aggregate relationship. The two figures on the right illustrate the same relationship from a temporal perspective for an individual state. The user chooses which state to visualize in the figures on the right by hovering over a dot within the left figure.}
    \label{fig:covaxxy_page}
\end{figure*}

We are not aware of any tools that concurrently explore the relationships between COVID-19 vaccine conversations, vaccine uptake, and epidemic trends. Consequently, we have created a web-based dashboard to fill this void.
The \textit{CoVaxxy} dashboard will track and quantify credible information and misinformation narratives over time, as well as their sources and related popular keywords.\footnote{\url{https://osome.iu.edu/tools/covaxxy}} 
Although we collect English tweets related to vaccines globally, the dashboard provides state-level statistics in the United States. Additionally, it shows global hashtag and domain sharing trends. 
It is updated daily. 
Figure~\ref{fig:covaxxy_page} illustrates one example of an interactive visualization that lets users visualize the relationship between various misinformation-related and COVID-19 pandemic data.
This data will be displayed alongside COVID-19 pandemic and vaccine trends. 
By highlighting the connection between misinformation and public health actions and outcomes, we hope to encourage the public to be more vigilant about the information they consume from their daily social media feeds in the fight against COVID-19.

\section{Data Characterization}

\begin{table}
\centering
\resizebox{\linewidth}{!}{%
\begin{tabular}{|l|l|l|l|}
\hline
\textbf{Users} & \textbf{Tweets} & \textbf{Hashtags} & \textbf{URLs} \\ \hline
1,847,067 & 4,768,204 & 39,857 & 983,158 \\ \hline
\end{tabular}%
}
\caption{Breakdown of the data collected between January 4 and January 11, 2021 in terms of unique users, tweets, hashtags, and URLs.}
\label{table:breakdown}
\end{table}

Our system started to gather tweets on Jan. 4, 2021. 
Table~\ref{table:breakdown} provides a breakdown of the dataset (as of January 11) in terms of the number of unique users, number of tweets they shared, and numbers of unique hashtags and URLs contained in these tweets.
Next let us analyze the data from that week to illustrate how our dataset might be used for different research projects.

\subsection{Volume} 

\begin{figure}[!t]
    \centering
    \includegraphics[width=\linewidth]{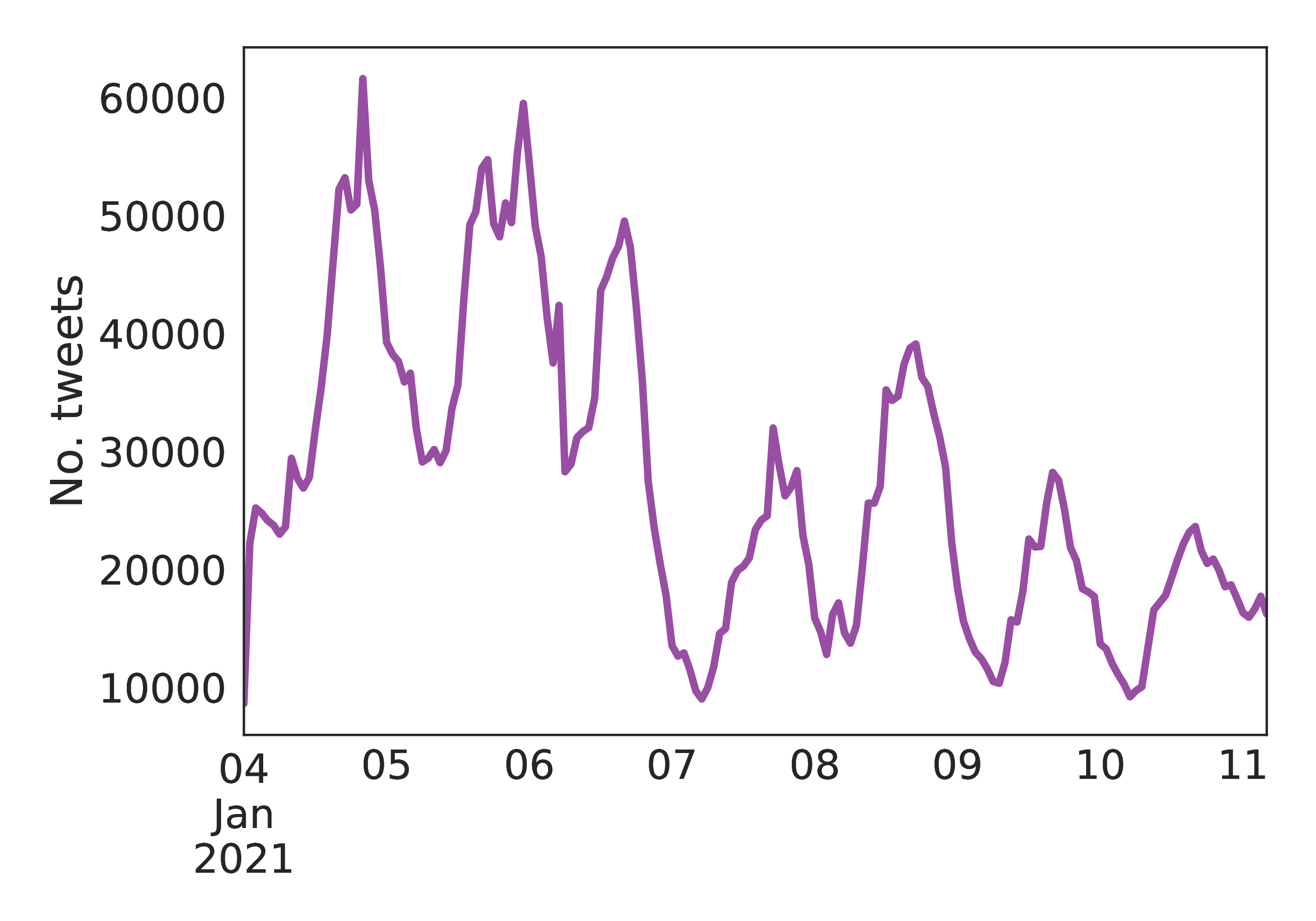}
    \caption{Number of collected tweets on an hourly basis since the beginning of the collection.}
    \label{fig:tweet_ts}
\end{figure}

We show in Figure~\ref{fig:tweet_ts} a time series for the number of tweets collected in our dataset, on an hourly basis. We can notice a decrease in the number of tweets after January 6, which might be driven by the increased media attention surrounding the storming of the U.S. Capitol.\footnote{\url{https://www.nytimes.com/2021/01/06/us/politics/protesters-storm-capitol-hill-building.html}} In fact, the mean daily number of tweets decreases from 900k tweets in the period of Jan 4--6 to 400k tweets in the period of Jan 7--11.

\begin{figure}[!t]
    \centering
    \includegraphics[width=\linewidth]{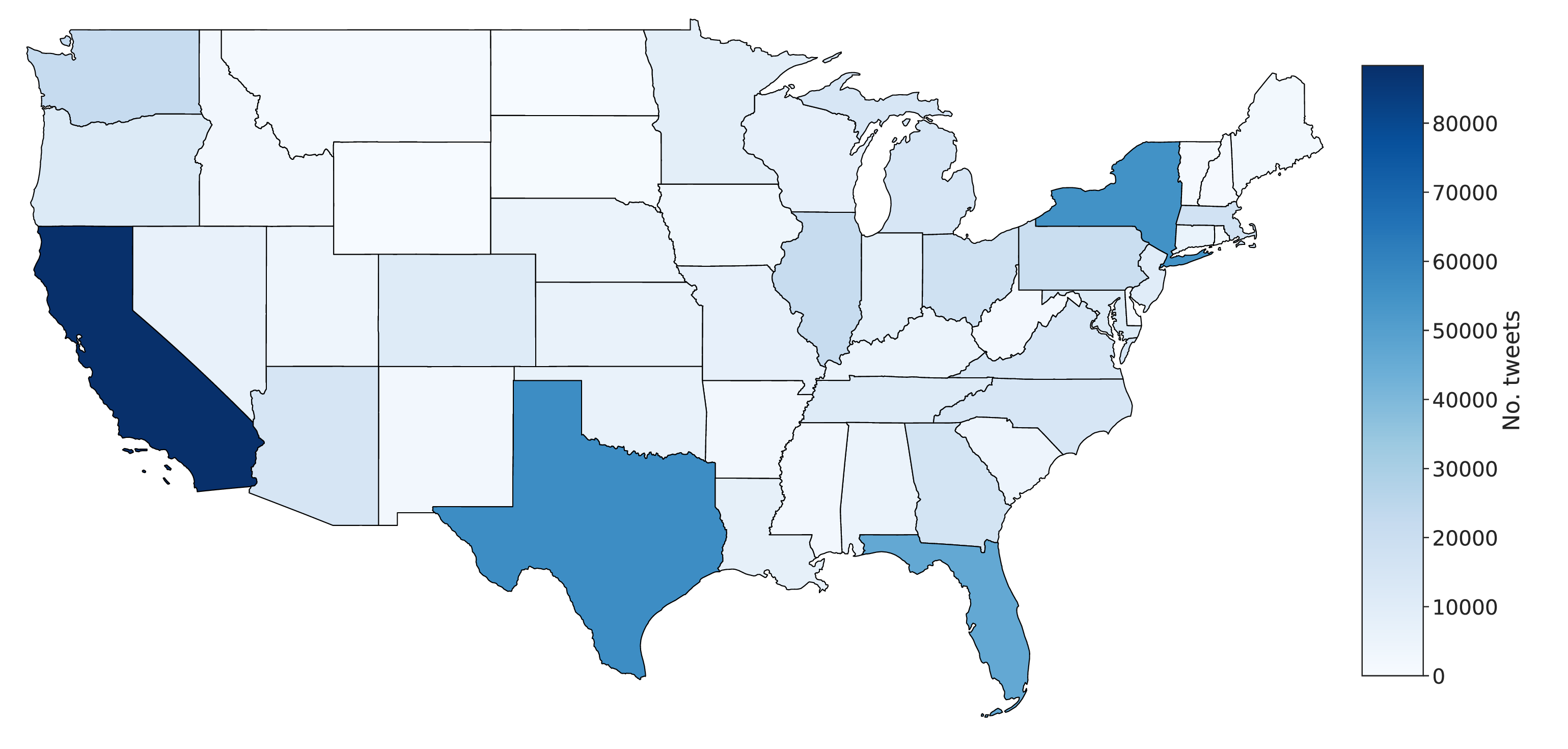}
    \includegraphics[width=\linewidth]{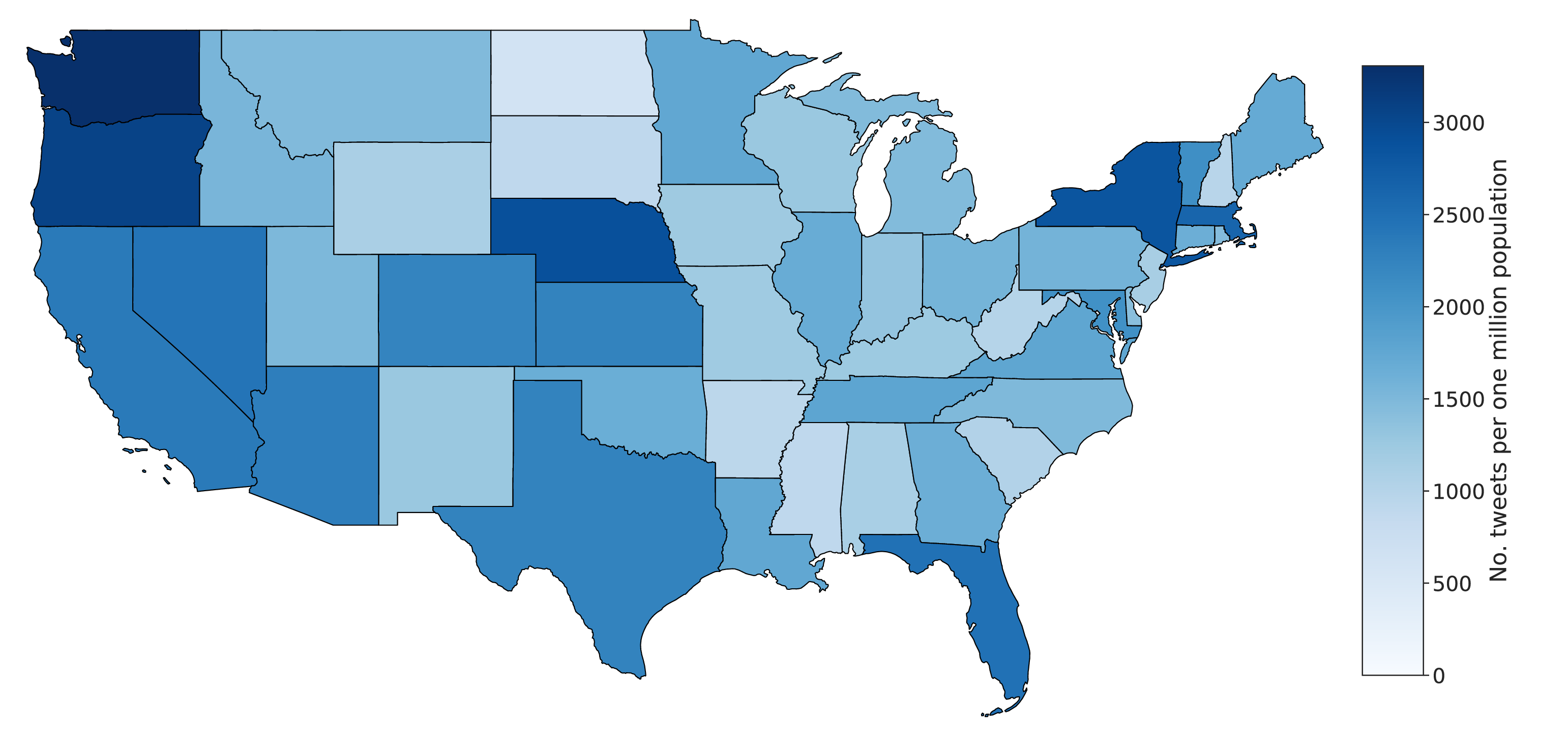}
    \caption{Distribution of the number of inferred geo-located tweets per U.S. state (excluding Alaska and Hawaii) by absolute numbers (top) and normalized by 2010 state population (bottom).}
    \label{fig:us_tw}
\end{figure}

In Figure~\ref{fig:us_tw} we show the distribution of the tweets geo-located in the contiguous United States. We use a naive approach to match tweets to U.S. states: we first extract the user location from the profile (if present) and then match it against a dictionary of U.S. states. Finally, we compute the number of tweets for each state based on the activity of users geo-located in that state. Over 1M users in our dataset have location metadata in their profile; we were able to match  approximately 40k users resulting in 600k geo-located tweets.
Providing an accurate methodology to geo-locate users is outside the scope of this paper; the reader should consider these results only as an illustration of the insights that can be gained from the \textit{CoVaxxy} data. 

\subsection{Hashtags}

\begin{figure}[!t]
    \centering
    \includegraphics[width=\linewidth]{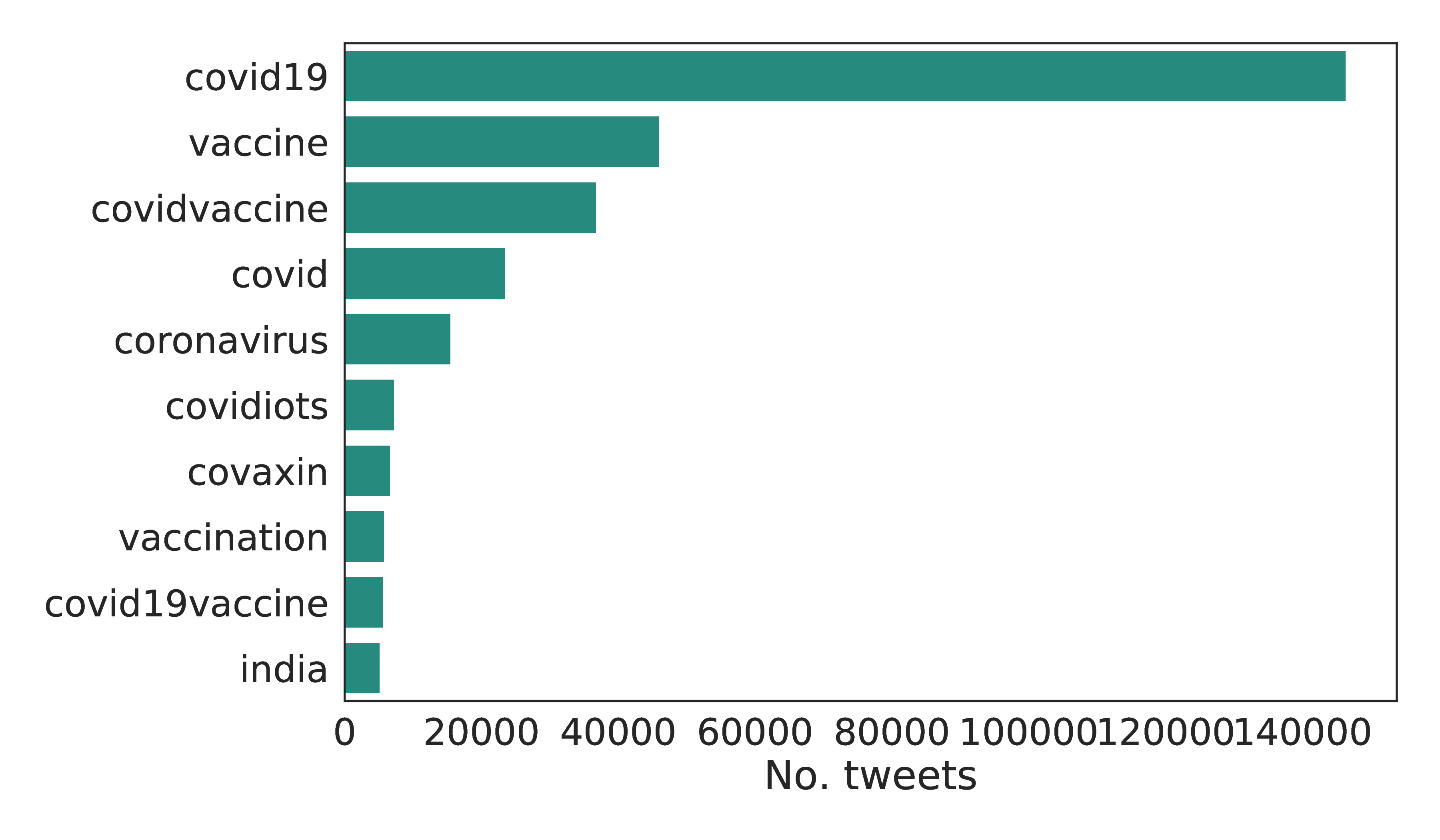}
    \caption{Top-10 shared hashtags.}
    \label{fig:hashtags_top10}
\end{figure}

Figure~\ref{fig:hashtags_top10} lists the most tweeted hashtags between January 4 and 11. We can see that they are largely related to the SARS-CoV-2 vaccine, with one (``\#covidiots'') referring to COVID-19 deniers. 

\begin{figure*}[t!]
    \centering
    \includegraphics[width=.7\textwidth]{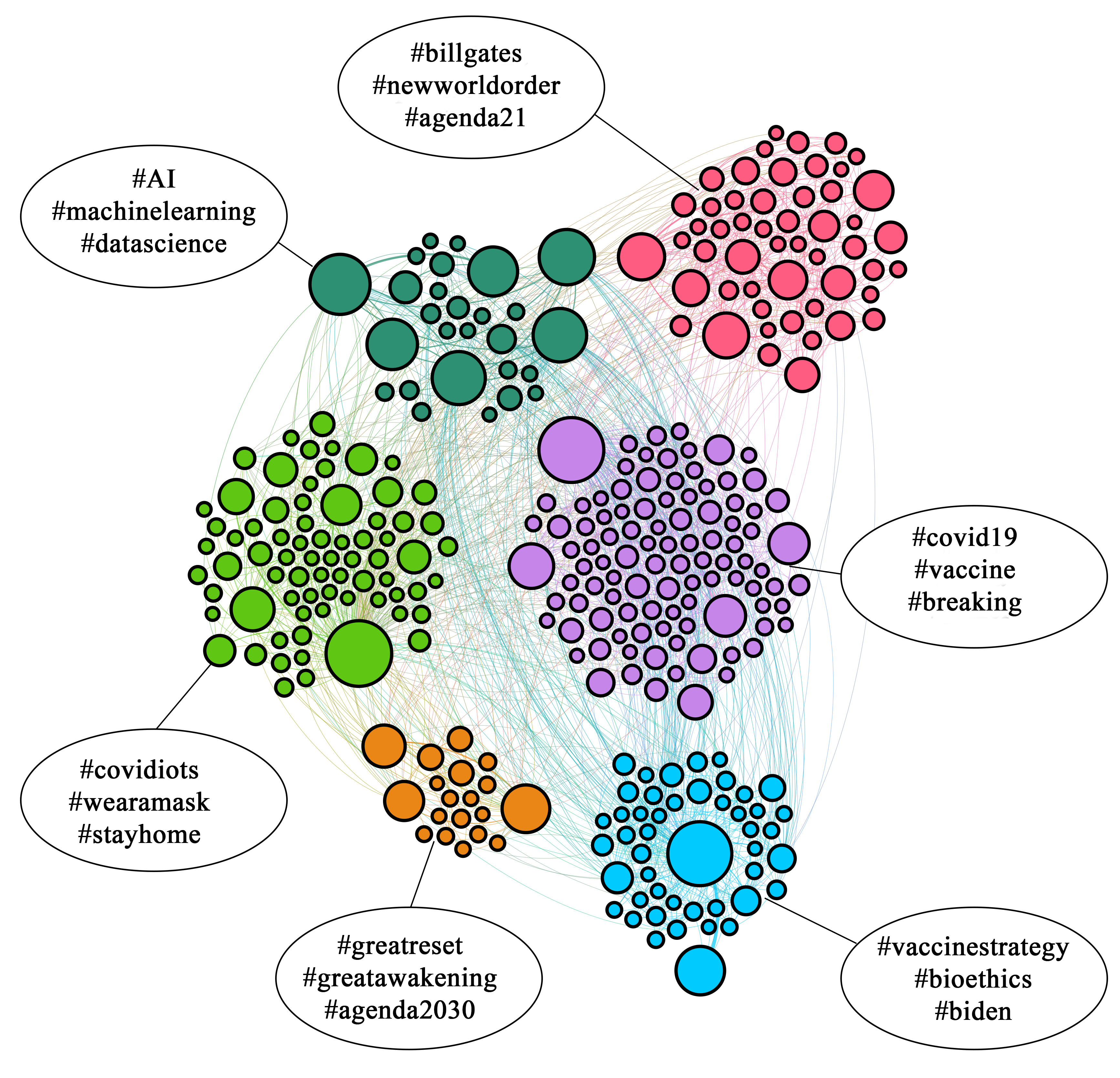}
    \caption{An overview of the prominent hashtags in the data, clustered into topic groups. A few hashtags characterizing each cluster are shown.}
    \label{fig:hashtags_groups}
\end{figure*}

Many different conversations can occur concurrently on Twitter, using different hashtags for different topics. To cluster related hashtags, we have grouped them together using a network algorithm. We form a co-occurrence network with hashtags as nodes and edges weighted according to how often the linked hashtags co-occur within tweets. Nodes are clustered using the Louvain method~\cite{blondelFastUnfoldingCommunities2008}. Groups with hashtags that are used the most are plotted in Figure~\ref{fig:hashtags_groups}. We observe  groups of hashtags associated with vaccine conspiracy theories (``\#greatreset,'' ``\#billgates'') as well as positive messages (``\#stayhome'').

\subsection{Sources}

\begin{figure}[!t]
    \centering
    \includegraphics[width=\linewidth]{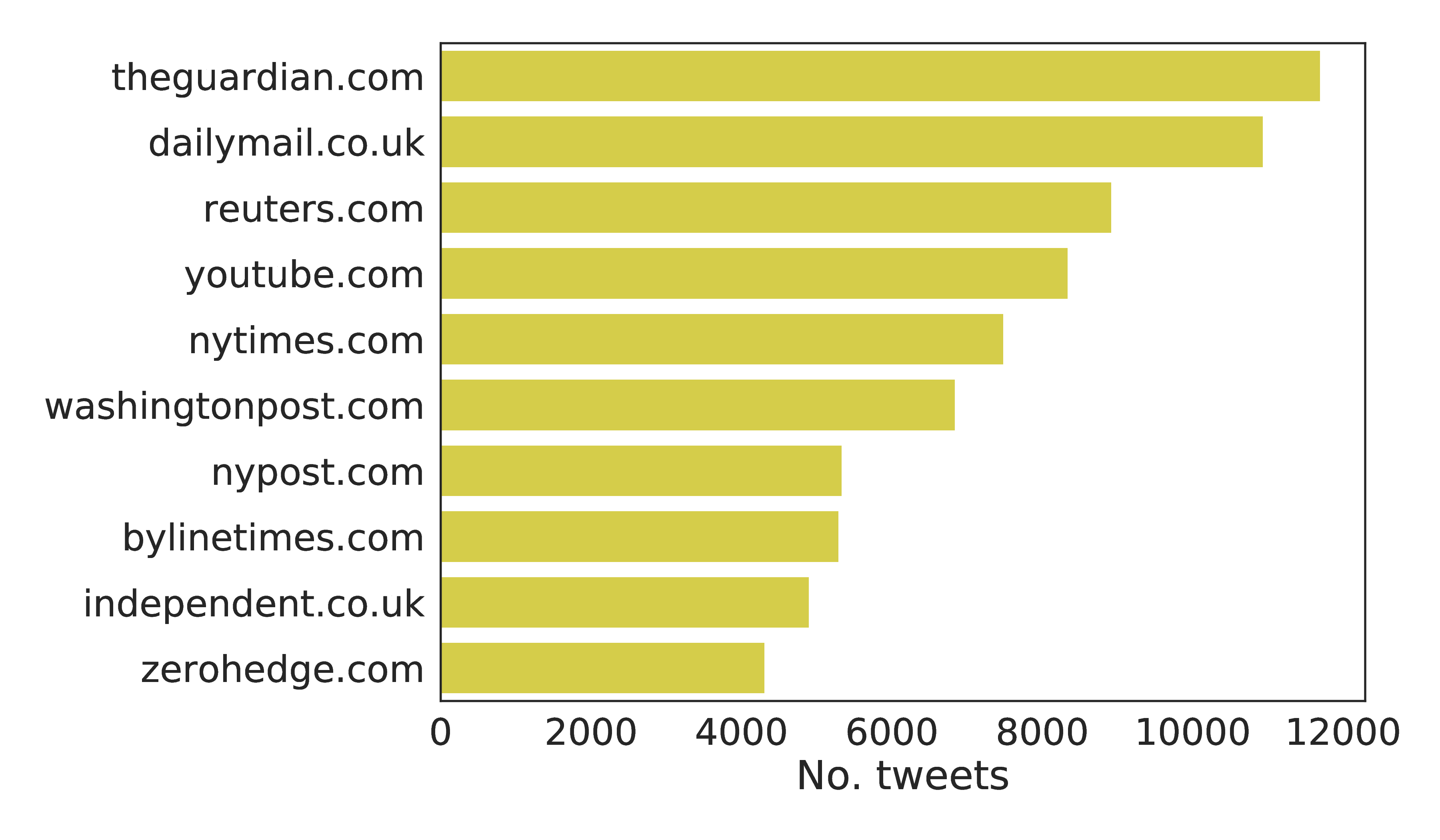}
    \caption{Top-10 sources shared in vaccine-related tweets.}
    \label{fig:domains_top10}
\end{figure}

In Figure~\ref{fig:domains_top10} we show the top-10 most shared websites. We exclude ``twitter.com,'' which accounts for over 3M tweets. These sites are comprised mostly of high-credibility information sources. However, one low-credibility source --- ``zerohedge.com'' --- also makes this list (see below for details on the classification). We also observe a large number of links to YouTube, which suggests further investigation will be needed to assess the nature of this shared content.

\begin{figure}[!t]
    \centering
\includegraphics[width=\linewidth]{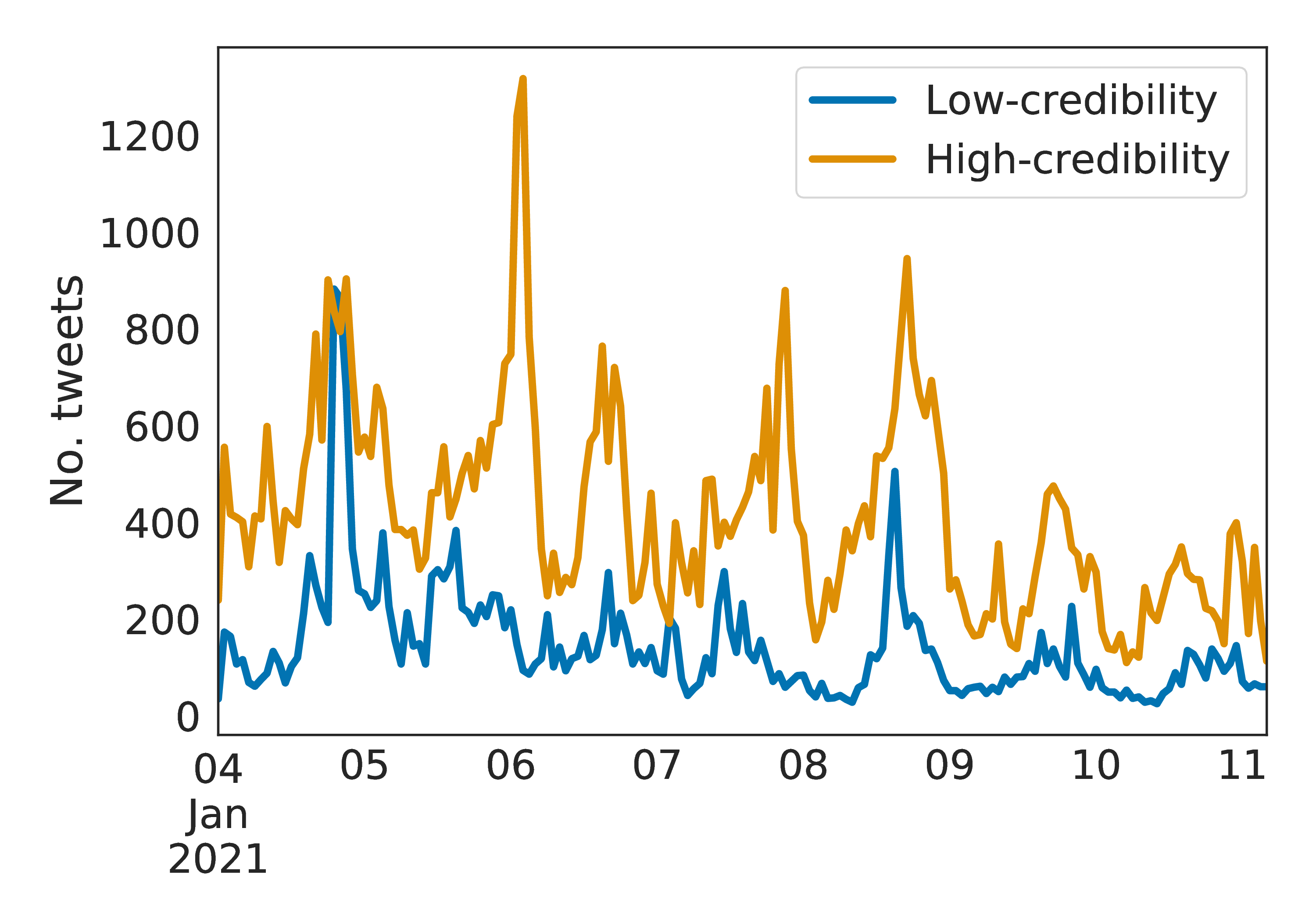}
    \caption{Number of hourly tweets containing links to low- (blue) and high-credibility (orange) sources.}
    \label{fig:low_high_tweet_ts}
\end{figure}

Figure~\ref{fig:low_high_tweet_ts} provides time series data illustrating the prevalence of low- and high-credibility information. We follow an approach widely adopted in the literature~\cite{Lazer-fake-news-2018, shao2018spread, Bovet2019, Grinberg, yang2020covid} to label links to news articles based on source reliability. In particular, we use a third-party list of 675 low-credibility sources\footnote{\url{https://iffy.news/iffy-plus/} (accessed November 2020)} and 26 hand-selected mainstream sources. 
The mainstream sources in this list are labeled by the Media Bias / Fact Check organization as having a factual reporting record as ``very high'', ``high'', ``mostly factual'' or ``mixed.'' We refer to them as ``high-credibility'' throughout the paper for simplicity. 
Overall, links to low-credibility sources account for 24,841 tweets compared to 72,680 tweets linking to our sample of mainstream sources. 
Readers should note that these numbers do not fully capture the news circulating on Twitter, as the lists we employ cannot be exhaustive.

\begin{figure}[!t]
    \centering
    \includegraphics[width=\linewidth]{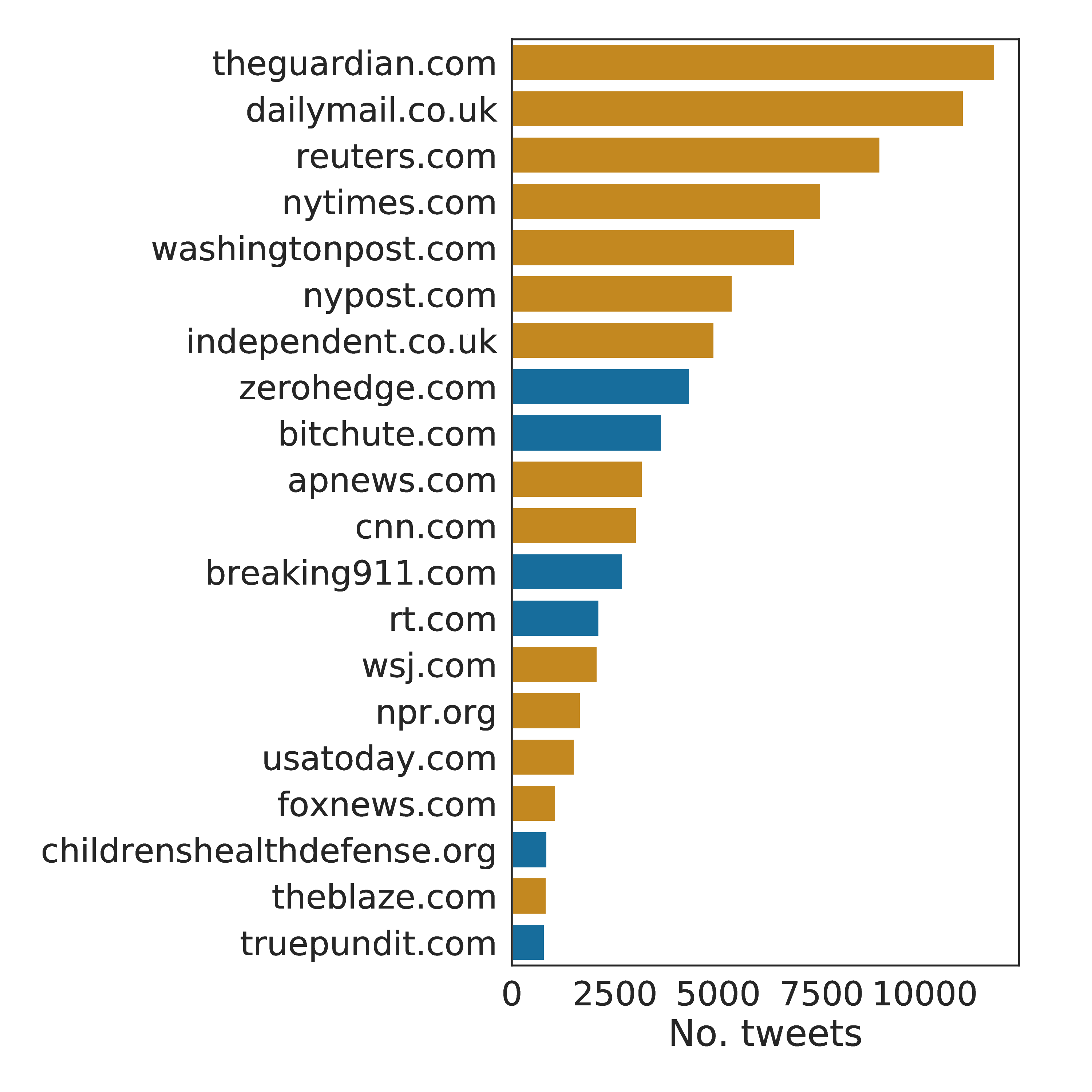}
    \caption{Top-20 shared low- (blue) and high-credibility (orange) sources.}
    \label{fig:low_high_top10}
\end{figure}

We further list in Figure~\ref{fig:low_high_top10} the 20 most shared news sources in both classes. We notice several unreliable sources (cf. ``zerohedge.com'' and ``bitchute.com'') that exhibit prevalence comparable to more reliable websites.  

\section{Discussion}

In this paper we present a new public dataset tracking discourse about COVID-19 vaccines on Twitter.
We characterize the data in several ways, including prominent keywords, geographic distribution of tweets, and clusters of related hashtags. We also present a data dashboard that visualizes statistics and insights from this data. 

In future work, we intend to explore the relationship between online discussion of COVID-19 vaccines and public health outcomes, like COVID-19 mortality and vaccine uptake. We will also leverage existing social media analysis tools to track emerging narratives and suspicious accounts, such as bots, coordinated campaigns, and troll farms~\cite{yang2019bot,yang2020covid, Pierri2020epj, Pierri2020scirep,pacheco2020uncovering}. Finally, we plan to explore models to better understand how vaccine misinformation and anti-vaccine sentiment spreads on social media.

This dataset has some key limitations. 
Critically, Twitter users are not a representative sample of the population, nor are their posts a representative sample of public opinions~\cite{pew2020twitterUsers}. 
Additionally, filtering our stream to include only English-language tweets comes at the price of occasionally excluding some variants of this language. This is because our stream gathers tweets that have been marked as containing English by Twitter's automatic language identification system, which may not capture some tweets by minority dialect speakers and multilingual speakers~\cite{jurgens-etal-2017-incorporating}.

The Twitter Filtered Stream API imposes a rate limitation of 1\% of all public tweets, which could limit our ability to capture all relevant content in the future. Fortunately, if this happens, Twitter provides the number of tweets not delivered within our stream. During the week described herein, we did not encounter this limitation.

Another potential source of bias is the keyword sampling procedure used to identify and collect COVID-19 vaccine related content, which involved evaluation of keywords to determine what was relevant. We are unable to fully exclude irrelevant content using only keyword-based filtering. However, further filtering is possible at a later stage. Other researchers may also refine the data to properly address their own topics of interest.

Given the large-scale, real-time nature of our data collection infrastructure, users do not have the ability to opt-out. This raises important ethical concerns related to anonymity. To address this concern, we note that (1)~our dashboard only displays aggregate data, obfuscating the ability of users to identify those captured within our data; and (2)~should a user delete a tweet or account, the related information will not be returned by Twitter during the re-hydration process.

The long-term aim of this project is to tackle the ambitious challenge of linking social media observations directly to  public health. We hope that researchers will be able to leverage the \textit{CoVaxxy} dataset to obtain a clearer picture of how vaccine hesitancy and misinformation affect health outcomes. In turn, such insight might enable public health officials to design better strategies for confronting vaccine hesitancy and refusal.

\section{Acknowledgments}

This work was supported in part by the Knight Foundation, Craig Newmark Philanthropies, DARPA (grant W911NF-17-C-0094), EU H2020 (grant 101016233 ``PERISCOPE''), and NSF (NRT award 1735095). Any opinions, findings, and conclusions or recommendations expressed in this paper are those of the authors and do not necessarily reflect the views of the funding agencies. 

\bibliography{bib}

\end{document}